\documentstyle[12pt,fps]{article}

\makeatletter
\@addtoreset{equation}{section}
\makeatother

\newcommand{\bibi}{\bibitem}

\newcommand{\eq}{\ref}
\newcommand{\beq}{\begin{equation}}
\newcommand{\eeq}{\end{equation}}
\newcommand{\bea}{\begin{eqnarray}}
\newcommand{\eea}{\end{eqnarray}}

\newcommand{\cc}{\cite}
\newcommand{\lb}{\label}

\newcommand{\ul}{\underline}
\newcommand{\lsim}{\stackrel{<}{\sim}} 

\newcommand{\al}{\alpha}
\newcommand{\bt}{\beta}
\newcommand{\lag}{\langle}
\newcommand{\rag}{\rangle}

\newcommand{\kp}{\kappa}
\newcommand{\lm}{\lambda}

\newcommand{\sg}{\sigma}

\newcommand{\ph}{\phi}

\newcommand{\ps}{\psi}

\newcommand{\psb}{\overline{\ps}}

\newcommand{\dsl}{\partial \!\!\!/}

\newcommand{\ra}{\rightarrow}

\newcommand{\be}{\begin{equation}}
\newcommand{\ee}{\end{equation}}

\def \3{\ss}

\newcommand{\Yu}{Y$_2$\ }

\def\dateandnumber(#1)#2#3#4{
\vbox to 18mm{%
     \hbox to \textwidth{ \hspace*{14mm} \hsize=40mm%
            \vbox{%
                 \hbox to 40mm{\large #1 \hss}%
                 \hbox to 40mm{    \hss}%
                 \hbox to 40mm{    \hss}%
                 }%
                 \hss \hsize=80mm%
            \vbox{%
                 \hbox to 80mm{\hss \large #2}
                 \hbox to 80mm{\hss \large #3}
                 \hbox to 80mm{\hss \large #4}
                 }%
            \hspace*{14mm} }%
      \vss
    }
}
\def\titleofpreprint#1#2#3#4{{\LARGE \bf
\vbox to 43mm{%
     \vss
     \hbox to \textwidth{ \hspace*{14mm} \hsize=130mm%
            \hss \vbox{
                      \hbox to 130mm{\hss \LARGE \bf #1\hss}%
                      \hbox to 130mm{\hss \LARGE \bf #2\hss}%
                      \hbox to 130mm{\hss \LARGE \bf #3\hss}%
                      \hbox to 130mm{\hss \LARGE \bf #4\hss}%
                 }%
            \hss \hspace*{14mm} }%
      \vss
    }}
}
\def\listofauthors#1#2#3{{\large
\vbox to 22mm{%
     \vss
     \hbox to \textwidth{ \hspace*{14mm} \hsize=130mm%
            \hss \vbox{
                      \hbox to 130mm{\hss \large #1\hss}%
                      \hbox to 130mm{\hss \large #2\hss}%
                      \hbox to 130mm{\hss \large #3\hss}%
                 }%
            \hss \hspace*{14mm} }%
      \vss
    }}
}
\def\listofaddresses#1#2#3#4{{\small
   \vbox to 18mm{%
        \vss
        \hbox to \textwidth{ \hspace*{14mm} \hsize=130mm%
               \hss \vbox{
                         \hbox to 130mm{\hss \small #1\hss}%
                         \hbox to 130mm{\hss \small #2\hss}%
                         \hbox to 130mm{\hss \small #3\hss}%
                         \hbox to 130mm{\hss \small #4\hss}%
                 	 }%
               \hss \hspace*{14mm}
        }%
        \vss
   }}
}
\def\abstractofpreprint#1{{\normalsize
\vbox to 110mm{%
     \vss
     \hbox to \textwidth{\hss \normalsize \bf Abstract \hss}%
     \normalsize
     #1
     \vss
     }}
}
\def\footnoteoftitle#1{{\small
\vbox to 30mm{\parindent0pt
     \vss\small #1 \vss
    }}
}
\def\footnoteitem(#1)#2{
\begin{list}{#1}{\labelwidth4.0mm \leftmargin7.0mm
\labelsep2.5mm \rightmargin7.0mm \parsep0.5ex plus0.2ex minus0.1ex
\itemsep0ex plus0.2ex }
\item #2
\end{list}
}
\begin{document}
\dateandnumber(February 1993     )%
{Oxford, OUTP-93-01P}%
{J\"ulich, HLRZ 93-15    }%
{                    }%
\titleofpreprint%
{           Study of the asymptotic freedom                    }%
{           of 2d Yukawa models on the lattice$^*$                 }%
{                                                              }%
{                                                              }%
{                                                              }%
\listofauthors%
{    A.K.~De$^1$, E.~Focht$^{2,3}$, W.~Franzki$^{2,3}$,        }%
{       J.~Jers\'ak$^{2,3}$ and M.A.~Stephanov$^4$             }%
{                                                              }%
\listofaddresses%
{\em \mbox{}$^1$Department of Physics, Washington University,  %
        St. Louis, MO 63130, USA                                }%
{\em \mbox{}$^2$Institute of Theoretical Physics E,             %
     RWTH Aachen,         D-5100 Aachen, Germany                 }%
{\em \mbox{}$^3$HLRZ c/o KFA J\"ulich,                          %
     P.O. Box 1913, D-5170 J\"ulich, Germany                    }%
{\em \mbox{}$^4$ Theoretical Physics, 1 Keble Rd.,            %
            Oxford OX1 3NP, UK                                 }%
\abstractofpreprint{
We investigate on the lattice the Yukawa models in 2 dimensions with
Z(2) and U(1) symmetries.  These models reduce to the usual and chiral
Gross-Neveu models, respectively, when the kinetic and the
selfcoupling terms of the scalar field are turned off.
The numerical data and mean field arguments suggest that, at least
for some range of the scalar field hopping parameter, fermion mass is
dynamically generated for arbitrarily weak Yukawa coupling. The
models are asymptotically free in this coupling, like the Gross-Neveu
models, even when the scalar quartic selfcoupling is strong.
}
\footnoteoftitle{
\footnoteitem($^*$){ \sloppy
Supported
by the US Department of Energy grant No. DE2FG02-91ER40628,
by Deutsches Bundesministerium
f\"ur Forschung und Technologie, by Deutsche Forschungsgemeinschaft
and by Jesus College, Oxford.
}
}
\pagebreak


\section{Yukawa models in 2 dimensions}

The 2d Yukawa models (Y$_2$)
have received little attention since the rigorous establishment
of some of their fundamental field theoretical properties in
the seventies (for a summary see ref.~\cc{GlJa87}).
The superrenormalizability of the Yukawa coupling
and the arbitrariness of the selfcoupling terms
of the dimensionless scalar field in 2d
might have led to a feeling that the \Yu models
have little relevance for the 4d field theories.
This is to be compared with the continuing interest in
the 2d Gross-Neveu (GN) and nonlinear $\sg$ (NL$\sg$) model
investigations, shown  e.g.\ by the recent exact mass gap
calculations in these models \cc{HaMa90,FoNi91}.

However, the \Yu models can be chosen so that
both the GN and NL$\sg$ models
are their special cases.
This is most obvious on the lattice.
For example, the action of the
Z(2) symmetric \Yu model on the lattice
can be chosen in the form
\bea
S & = & -2\kp\sum_{x,\mu}\ph_x\ph_{x+\mu}+\sum_x\ph_x^2+
          \lm\sum_x(\ph_x^2-1)^2 \nonumber \\
& + & \sum_{x,\al} \psb_x^\al\dsl\ps_x^\al
      + y\sum_{x,\al}\psb_x^\al\ph_x\ps_x^\al~.
\lb{ACTION}
\eea
Here we have introduced $N$ ``naive'' Dirac fermion fields
$\psi^\al$, $\al = 1,$...,$N$, on the lattice
which, due to the fermion doubling, describe $N_F=4N$ Dirac
fermions  of zero bare mass.
All the fields and the couplings are made dimensionless by the
appropriate rescaling with
the lattice constant $a$, $x$ enumerates lattice sites
and $\partial_\mu$ is the lattice derivative.
The action for the U(1) symmetric \Yu model which we study as well
has the form similar to (\ref{ACTION})  with two-component
field $\vec\phi_x = (\phi^{1}_{x}, \phi^{2}_{x}) $ instead of
$\phi_{x}$ and the Yukawa term
\mbox{$y\sum_{x,\al}\psb_x^\al(\phi^{1}_{x} +
i \phi^{2}_{x}\gamma_P)\ps_x^\al$}
(where $\gamma_P$ is the 2d analog of $\gamma_5$).

At $\kp = \lm = 0$ the action (\eq{ACTION}) describes the
Z(2)-symmetric GN model in the auxiliary scalar field representation
of the 4-fermion coupling.
The Yukawa coupling $y$ is related to the usual GN coupling $g$ by
$y = \sqrt{2}g$.
On the other hand, at $\lm = \infty$ and $y = 0$ the action (\eq{ACTION})
describes the Ising model.
The U(1) symmetric \Yu model in similar cases reduces to the chiral GN
model or to the XY model, i.e.\ the NL$\sg$ model with U(1) symmetry,
and similarly for other symmetry groups.
The \Yu models thus interpolate between the GN and spin or NL$\sg$
models.

On the lattice,
by choosing the above formulation of the scalar
field sector using the hopping parameter $\kp$,
the kinetic term can be turned on or off {\em gradually},
illustrating the smoothness of the transition from an auxiliary
to a dynamical scalar field.
As the numerical study of lattice Yukawa models in 4d
showed, the physical observables behave continuously with $\kappa$
in the vicinity
of $\kappa=0$ including a region of negative $\kappa$
\cc{HaHa91,BoDe92b,DeJe92}.
This fact elucidates the relation between the Yukawa and four-fermion
theories found in the continuum in the leading order in $1/N_F$
expansion \cc{HaHa91,Zi91}.

Motivated by these considerations and by
the recent discussion of a relationship between
the Nambu--Jona-Lasinio type four-fermion theories and the Standard
Model \cc{HaHa91,Zi91} (for a recent review see ref.~\cc{Ba93}),
we address here the question to what extent
the Z(2) and U(1) \Yu
models still possess the most interesting and important properties of
the GN models \cc{GrNe74,Wi78b},
namely the asymptotic freedom of the Yukawa coupling
$y$, the    dynamical fermion mass generation and, in the case of the
Z(2) model, the    dynamical symmetry breaking.


\section{Expected scaling properties}

In the GN models the basic scaling properties at $y \ra 0$ can be
derived by means of the 1/$N_F$ expansion.
For \Yu models this expansion is applicable only for \ul{small $\lm$},
strictly speaking for~$\lm$ = O(1/$N_F$) \cc{HaHa91,Zi91}.
It gives results very similar to those for the GN models, provided
$\kp < \kp_c(\lm)$.
Here $\kp_c(\lm)$ is the critical line
of the scalar model at $y=0$ (i.e.\ without fermions)
 which for $-\kp_c(\lm) < \kp < \kp_c(\lm)$ is in the high-temperature
phase.
The fermion mass
 in lattice units $am_F$ is expected at fixed $\kp$ and $\lm$
to scale with $y$ as
\beq
am_F \propto \exp\left[-\frac{1}{2\bt_0}
\frac{a^2m_\phi^2}{Z_{\phi}}\frac{1}{y^2} \right].
\lb{ASY}
\eeq
Here $\bt_0$ is the first coefficient of the $\bt$-function of
the continuum GN model with $N_F$ flavours of Dirac fermions,
\beq
\bt_0=\frac{N_F-1}{2\pi} \quad(\mbox{Z(2)}),\qquad
\bt_0=\frac{N_F}{2\pi} \quad(\mbox{U(1)}),
\lb{BT0}
\eeq
 $am_\phi$ is the $\ph$-field mass
 in lattice units and $Z_\phi$
its renormalization constant at $y=0$.
For $\lm=0$ we have    at $y=0$ the free scalar field theory with
\beq
      a^2 m_\phi^2 =  \left(1-\frac{\kp}{\kp_c(0)}\right)
      \frac{1}{\kp}
      ~,~~ Z_\phi = \frac{1}{2\kp}~,
\lb{M}
\eeq
%
%
   and $\kp_c(\lm=0) = 1/4$.
Note that for small $\lm$
the ratio $Z_\phi/a^2m_\phi^2$ appearing in eq.~(\eq{ASY})
is the scalar field propagator
at zero momentum, i.e.\ the susceptibility $\chi$
of the pure scalar model at the same values of $\lm$ and $\kp$.
Although $a^2m_\phi^2$ and $Z_\phi$ are not well defined separately at
$\kp\le0$,
the ratio $\chi$ is.
Thus the scaling law (\eq{ASY}) is sensible also at $\kp \le 0$.
We shall see below that for small $\lm$ the  mean field
approximation (MFA) gives the same results as the $1/N_F$
expansion.

For \ul{large $\lm$} the 1/$N_F$ expansion is a priori not applicable.
However, as we shall argue in the next section,
 the MFA can be applied to study
the \Yu models at all $\lambda \ge 0$.
 The effective interaction produced by fermions is
very nonlocal and favours ferromagnetic ordering of the field
$\phi_x$. Such an interaction can be well described by some effective
mean field $H$ acting on $\phi_x$ at each site.
Thus using the MFA we reduce the model to a pure scalar model with only
local interactions and the external field $H$.
Such a model can be easily studied
e.g.\ by high temperature expansion or Monte Carlo (MC) simulation.

The results we obtain using this approach are as follows:

{\it (i)}~~~ At any nonnegative $\lambda$ including
$\lambda=\infty$ the \Yu  model with Z(2) symmetry is in the
phase with broken symmetry ($\langle \phi \rangle \neq 0$) for
arbitrarily small $y$.
Similarly, the corresponding U(1) model is in the spin wave phase
(analogous to the low temperature phase of the $XY$ model).
In both cases fermion mass is  dynamically generated for arbitrarily
small $y$ and vanishes only at $y=0$.
This means that the Yukawa coupling is asymptotically free also for
 any nonnegative $\lm$.

{\it (ii)}~~~ For $\kp < \kp_c(\lm)$ the fermion mass and,
in the Z(2) case also the magnetization
$y \langle \phi \rangle$, scale with $y$ according to
\beq
am_F \propto \exp\left[-\frac{h(\kp,\lm)}{y^2} \right].
\lb{ASYh}
\eeq

{\it (iii)}~~ In the MFA the function $h(\kp,\lm)$ is given by
\beq
h(\kp,\lm)=\frac{\pi}{N_F}\frac{1}{\chi},
\lb{h}
\eeq
where $\chi$ is the zero-momentum scalar propagator in the pure scalar
$\phi^4$ model at given $\kappa$ and~$\lambda$. If the propagator is
dominated by a pole with mass $am_\phi$ and residue $Z_\phi$, then

\beq
          \chi \simeq \frac{Z_\phi}{a^2m_\phi^2}
\lb{CHI}
\eeq

Thus according to the MFA the asymptotic freedom and the other
mentioned properties of the GN models occur also for large
$\lm$ in the \Yu models.
Our Monte Carlo data at $\lm = 0.5$ and $\lm = \infty$
support {\it (i)} and {\it (ii)}
(see also ref.~\cc{DeFo93a}), whereas some deviations
from {\it (iii)} are observed.


\section{The mean field approximation}

Integrating over fermion variables in the partition function for the
Z(2) Yukawa model (\ref{ACTION}) we obtain an effective scalar
model with the contribution to the action from the fermion
determinant

\beq
S_{det}[\phi]=-Ntr\ln M[\phi];
\lb{Sdet}
\eeq
where
 \[ M[\phi]_{xz}=\frac{1}{2}\gamma_\mu(\delta_{z,x+\hat{\mu}} -
\delta_{z,x-\hat{\mu}}) + y\phi_x \delta_{xz} \equiv K_{xz} +
 y\phi_x \delta_{xz}.  \]
 We know from the
experience with Yukawa models in four dimensions
   (for a recent review see ref.~\cc{DeJe92})
that fermions
strengthen ferromagnetic ordering: the transition from disordered to
ordered phase occurs at smaller values of $\kappa$ as $y$ increases from
zero. As was pointed out to us by E.~Seiler \cite{Se92} one can prove
also that $S_{det}[\phi]$ is minimized on the totally ordered
configuration.

We first give a very straightforward derivation of the announced results
by the MFA method and then discuss its reliability.
   As infinitely many sites
participate in the interaction $S_{det}$ with a given one, say $x$, we
expect their effect on $\phi_x$ to average into an effective mean
field $H$ whose fluctuations are negligible.
 So we obtain a scalar $\phi^4$ model with only local interactions
given by
the first three terms in (\ref{ACTION}) to which the constant
external field $H$ is applied.
The mean magnetization $\sigma$ of such a model
is given  at each $\kappa$ and $\lambda$
by the response function $\sigma = f(H)$.

The simplest approach to calculate $H$ is to differentiate $S_{det}$
with respect to $\phi_x$ at the site $x$
and then substitute the mean value $\sigma$
for all $\phi$'s. This gives
\beq
H(\sigma) = 2Ny^2\sigma \int \frac{d^2p}{(2\pi)^2} \left(
\sum_\mu \sin^2 p_\mu  + (y\sigma)^2 \right)^{-1}.
\lb{Hsigma}
\eeq
Selfconsistency then requires that
\beq
\sigma=f(H(\sigma)),
\lb{self}
\eeq
where $H(\sigma)$ is from (\ref{Hsigma}). For given $y$, $\kappa$ and
$\lambda$ one can find $\sigma$ as a solution of this equation.

At $\kappa < \kappa_c(\lambda)$
for small $\sigma$ we can write
\beq
f(H)=\chi H + O(H^3),
\lb{chi}
\eeq
where $\chi$ is the susceptibility of the scalar $\phi^4$ model, and
(\ref{Hsigma}) gives
\beq
H(\sigma) = \frac{4}{\pi}Ny^2\sigma\ln\frac{1}{y\sigma} + O(\sigma).
\lb{log}
\eeq
Substituting (\ref{chi}) and (\ref{log}) into (\ref{self}) one finds
that a nonzero solution exists for arbitrarily
small $y$. Taking $am_F=y\sigma$ we
arrive at eqs.~(\ref{ASYh}) and (\ref{h}) with $N_F=4N$.
Only the slope~$\chi$ of the response function $f(H)$ at the origin
was needed to derive this asymptotic scaling law. As we shall see in
Sect.4, the eq.(\ref{self}) with the full response function can give a
very good approximation for $m_F$ even when $\sigma$ is not small.

Now let us discuss the MFA method in more detail. To get an idea of how
nonlocal the interaction $S_{det}$ is at small $y$ and whether it can
produce the effective mean field let us expand it in powers of $y$:

\beq
S_{det}[\phi]=const + N\frac{y^2}{2}\sum_{xz}Tr K^{-1}_{xz}
K^{-1}_{zx} \phi_x \phi_z + O(y^4),
\lb{expansion}
\eeq
where $Tr$ is the trace over Dirac indices. The term of order $y^2$
produces a nonlocal interaction between $\phi_x$ and
$\phi_z$:

\beq
S_2 = -Ny^2 \sum_{xz} J_{xz} \phi_x \phi_z
\lb{S2}
\eeq
with $J_{xz}=\frac{1}{2}Tr K^{-1}_{xz}K^{-1}_{xz}  $
( as $K^{-1}_{xz}=-K^{-1}_{zx}$). One can
find out easily that $J_{xz} > 0$
when $x$ and $z$ are separated
 by an odd number of links and $J_{xz} = 0$
otherwise. Thus this interaction is ferromagnetic.

The MFA works well when fluctuations of the
mean
field at site $x$, \mbox{$H_x=2Ny^2 \sum_{z} J_{xz} \phi_z$},
are small compared
to $\lag H_x\rag$. In 2 dimensions $J_{xz}$ falls off with $|x-z|$ so slowly
that it produces the infrared logarithmic divergence

\beq
\sum_z J_{xz} = \frac{1}{2} Tr K^{-2}_{xx} = \int \frac{d^2p}{(2\pi)^2}
 \left(\sum_\mu \sin^2 p_\mu \right)^{-1}.
\lb{sumJ}
\eeq
or, in other words, very large number of sites contributes effectively
to $H_x$. This means that the mean value of $H_x$ grows with the size
$L$ of the system as $Ny^2\langle\phi\rangle\ln L$ if there is some
nonzero $\langle \phi \rangle$. As one can check the fluctuations of
$H_x$ around its mean value are finite at $L\to\infty$ and thus at
large $L$ the MFA is justified if $S_2$ was    
the only interaction.
Moreover, the divergence in $\lag H_x\rag$ means that such
a system would be ordered at any $y$.  This can be also understood if
one considers the free energy of the long wavelength modes of $\phi$.
It is negative and diverges as $\ln L$.

The interaction $S_{det}$ contains also other nonlocal terms
   which are of higher order than~$y^2$. They
obviously contribute to $H$ in the expression (\ref{Hsigma}): the mass
term $am_F=y\sigma$ can be interpreted as their effect.
However, one can easily
realize that the interaction producing the mean field $H(\sigma)$ in
(\ref{Hsigma}) almost coincides with $S_2$ for the distances smaller
than $O(1/m_F)$ and falls off exponentially at larger distances.
This means that when $m_F$ becomes smaller the MFA works better as
more sites participate effectively in the interaction.
The size $1/m_F$ plays the role of the infrared cutoff which controls
the logarithmic divergence in eq.~(\ref{log}).

It is also very useful to
look at what happens from the point of view of the free energy of the
long wavelength modes, i.e.\ effective potential.
The effective potential $V_{eff}(\sigma)$ at the origin behaves like
$V_{eff}(\sigma) \sim y^2\sigma^2\ln y\sigma $ (this is clear from the fact
that $H(\sigma)$ in (\ref{Hsigma}) is minus the derivative of the fermion
contribution to the $V_{eff}(\sigma)$) and thus $\sigma=0$ is a local
maximum.
Approach based on the effective potential was applied in a recent
paper \cc{She93} and gives the same results as the MFA method, as one
would expect.
We think, however, that the MFA approach allows us to understand
better the nature and the effect of the infrared singularity, the
mechanism of the fermion mass generation in the U(1) case
(see below) and the subtleties of both methods.

Although our MFA method as well as the
effective potential approach \cite{She93} and their results might look
very consistent one
should be cautious and not overestimate their reliability. The
expression (\ref{Hsigma})    has a form of a
contribution of one loop with massive fermions. Higher loop
contributions may turn out to be important. For example, for the GN
model ($\kappa=\lambda=0$) the selfconsistency equation (\ref{self})
coincides with the gap equation obtained in the leading order in
$1/N_F$ expansion.
However, we know that in the Z(2) GN case the coefficient $h$ of
$1/y^2$ in the scaling law given by (\eq{h}) receives $1/N_F$
corrections (compare with eq.~(\eq{BT0})). The same result for $h$ as
eq.~(\eq{h}) is obtained in \cite{She93} and must receive finite
corrections for this reason as well. The terms in $V_{eff}$ of higher
order in $y^2$ neglected in \cc{She93} contain in fact powers of
$y^2\ln(1/y\sigma)$ and are not negligible as $y\to 0$.
Nevertheless, one can expect that $H$ still has the singularity similar
to (\ref{log}) at $m_F\to 0$ as a consequence of the strong
nonlocality of the interaction $S_{det}$ in $2$ dimensions.
One can also expect that $m_F \neq 0$ if $\sigma \neq 0$.
Then the conclusion {\it (i)}
that the fermion mass is generated for arbitrarily small
$y$ still holds.
It is natural to expect that these features do not depend on the
structure of the local scalar interactions in (\ref{ACTION}),
in particular on the value of $\lambda$.
The predictions {\it (ii)} and in particular {\it (iii)}
are less reliable, however.

Let us now see how one can apply the MFA in the \Yu model with the U(1)
symmetry, when the symmetry cannot be broken spontaneously.
It is known that this fact does not prevent
fermions from acquiring a mass \cite{Wi78b}.
To understand how this happens we consider the interaction of the
scalar field induced by the fermion with a small mass $m_F$ in one loop
that we have already discussed.
For distances between sites smaller
than $O(1/m_F)$ one can neglect the mass and the interaction behaves
like (\ref{S2}). Such interaction can produce ferromagnetic ordering
on the distances smaller than $O(1/m_F)$ if $m_F$ is small enough.
On the
distances larger than $O(1/m_F)$ long wavelength fluctuations (spin
waves) destroy the ordering as usual in two dimensions.
However, in a finite volume of the linear
size $O(1/m_F)$ the magnetization is nonzero and its direction is
drifting slowly.

Now let us imagine a fluctuating scalar field with such properties
playing the role of a background field coupled to a fermion via Yukawa
interaction. For small $y$ we expect the fermion to acquire a mass
given by $am_F\simeq y\lag\phi\rag$.
What $\lag\phi\rag$ enters this formula?
It is reasonable to expect that the fermion ``feels'' only the
ordering on the scale of its Compton wavelength,
so that $\langle \phi \rangle$ should be averaged over a volume
of the size $O(1/m_F)$.
This agrees with the observation made
by Witten that ``almost long-range order'' in chiral GN models is
sufficient to generate the fermion mass \cite{Wi78b}.

Bearing this in mind we can carry out the mean field
considerations in the U(1) case in
complete analogy with the Z(2) case.
One has to realize, however, that $\lag\phi\rag\equiv\sigma$ in the
formulas is the average magnetization on the scale of $O(1/m_F)$.
We actually see nonzero $\lag\phi\rag$ in a
finite volume in MC simulations which agrees with $am_F=y\langle \phi
\rangle$
remarkably well (within few percent) when $L\approx 1/am_F$.

We now compare the solutions of the equations (\ref{Hsigma}) and
(\ref{self}) on finite lattices with numerical results and use these
equations for data analysis.


\section{Numerical results}

We simulated the Z(2)-symmetric \Yu model defined by the action
(\eq{ACTION}) and the analogous model with the U(1) symmetry
on the lattices $L^2 = 16^2, 32^2$ and $64^2$, mostly with $N_F = 16$.
The expected scaling behaviour (\eq{ASYh})
suggests to collect data at many $y$-points at fixed
values of $\lm$ and $\kp$ and to study the $y$-dependence of the
fermion mass.
We have chosen several $\kp$ points in the interval
$-0.3 \le \kp < \kp_c(\lm)$
with $\lm = 0, 0.5$ for both Z(2) and U(1) models and
$\lm = \infty$ for the U(1) model%
\footnote{The hybrid MC algorithm for simulating dynamical fermions
we are using unfortunately does not work in the Z(2) case at
$\lm =\infty$
because of the discrete nature of the variables.}.
For each pair of $(\kp,\lm)$ values and $L$ we determined the fermion
mass $am_F$ at several $y$ values such that $2/L \lsim am_F \le 1$.

As has been reported in ref.~\cc{DeFo93a} in detail, the dependence
of $am_F$ on $y$ at all investigated $(\kp,\lm)$ points
is consistent with the expectations {\em (i)} and {\em (ii)}.
The data analysis on finite lattices has been performed
by means of the generalized gap equation
\be
\frac{h}{y^2}=\frac{\pi}{2L^2}
\sum\limits_{ \{p\} }
\frac{1}{\sum\limits_\mu\sin^2p_\mu+(am_F/s)^2}
\lb{GAP}
\ee
which gives the scaling law (\eq{ASYh}) at $L=\infty$
and describes     quite well the onset of finite size effects
and a very slow approach to the asymptotic scaling.
The values of the fit parameters $h$ and $s$
depend only on  $\kp$ and $\lm$ but not on $L$.
Note that eq.~(\ref{GAP}) at $s=1$ can be viewed as a
selfconsistency equation like (\ref{self}) with a linear response
function whose slope is a free parameter.

In the present paper we compare these results with the
parameter-free predictions of the MFA,
to see how well it can describe the data.
To obtain these predictions we solve for given $\kp$, $\lm$ and $y$
the finite lattice version of eq.~(\ref{Hsigma}) and eq.~(\ref{self})
numerically. On a finite lattice
we substitute the integral in (\ref{Hsigma}) by the discrete sum
over the momenta allowed by the boundary conditions on the fermion
fields: periodic in one and antiperiodic in another direction.
The response function $f(H)$ is obtained by a MC simulation of
the scalar $\phi^4$ model in the external field. For $\kp=0$ the
response function can be also expressed in terms of
easily computable one dimensional integrals.
So for each $(\kp,\lm)$ pair and $L$ we obtain a function
$am_F(y)$ which we compare with the data.

In the case of the Z(2) model we find only a qualitative agreement.
There is a systematic discrepancy between the data and
the MFA prediction at any $\kp$ and $\lm$ values.
This is
connected with the fact, discussed in the previous section, that in the
scaling limit $am_F\to 0$ the one-loop formula for
$H(\sigma)$ (\ref{Hsigma}) does not reproduce the correct value of
the $\beta_0$ coefficient in the scaling law (\ref{ASY}) already
for the GN model $(\kp=0, \lm=0)$.
We note that this happens in the Z(2) GN model also in the leading
order of the $1/N_F$ expansion.
It would be very useful to include the higher loop corrections but at
the moment we do not know how to do that at large $\lm$
systematically.\footnote{Some additional complications in the
subleading order in $1/N_F$ arise in the Z(2) model even at $\lm=0$
due to the naive fermion discretization in eq.~(\eq{ACTION}).}
Without these corrections the simple MFA that we used apparently cannot
describe the data in the Z(2) model sufficiently well
quantitatively.

In the  chiral GN model higher loop corrections do not modify the
coefficient of  $1/y^2$ in the exponent of the scaling law.
Correspondingly, the agreement is much better in the U(1) \Yu
models. For comparison we present plots for
$(\kp,\lm) = $ $(0,0)$, $(0.2,0)$, $(0,\infty)$ and $(0.2,\infty)$.
At $\lm=0$ the $1/N_F$ expansion can be applied and in the leading
order coincides with the MFA prediction.
The data agree with this prediction well (see figs.~1 and 2) as one can
expect if $N_F$ is large.
For $\kp=0$ at $\lm=0.5$  and $\lm=\infty$ (fig.~3) the data agree
with the MFA prediction as well as at $\lm=0$. However, if
$\kp\neq 0$ we observe both at $\lm=0.5$ and $\lm=\infty$ a significant
discrepancy illustrated for $\lm=\infty$ and $\kp=0.2$ in fig.~4.

We see that the MFA works not only at $\lm=0$ but also at large
$\lm$ as long as $\kp=0$ even better than one could naively expect.
The discrepancy at $\kappa\neq0$ and large $\lambda$ could be due to the
crudeness of the simple MFA that we used and more systematic study is
needed to understand it fully.
In any case, as the fits with the generalized gap equation (\eq{GAP})
performed in ref.~\cc{DeFo93a} demonstrate,
all our data are consistent with the scaling
of the general form (\eq{ASYh}).
Here we show such a fit in fig.~4 (dotted-dashed line).
The values of the fit parameters in eq.~(\eq{GAP}) are approximately
$s=0.27$ and $h=0.86 (\pi/N_F\chi)$.
We cannot draw firm conclusions from these numbers as they are
obtained in the region where the asymptotic scaling is not yet
achieved and we do not know how well the eq.(\ref{GAP}) describes the
approach to the asymptotic scaling.

In conclusion, both our numerical data and mean field considerations
 suggest that the  Z(2) and U(1) Yukawa models in 2d
have asymptotically free Yukawa coupling and exhibit dynamical
fermion mass generation also for arbitrarily strong scalar field
selfcoupling, when the usual perturbative and $1/N_F$ expansions
are not applicable.
They are thus quite similar to the corresponding
Gross-Neveu models.
However, at present we cannot say whether they belong to the same
universality class, i.e.\ what  the precise form of the scaling law~is.
Our precision is also not yet good enough to compare with the
results of the mass gap calculations~\cc{FoNi91}.
Further study of these questions, as well as of the transition
from the Gross-Neveu-like behaviour to the NL$\sg$ or
spin model cases is needed.
Finally, we have not yet studied the properties of the 2d lattice
Yukawa models at larger negative $\kp$, where the antiferromagnetic
scalar field coupling competes with the
Yukawa coupling which supports ferromagnetic order.

\subsection*{Acknowledgements}

We thank E.~Seiler and M.M.~Tsypin for helpful suggestions,
A.~Hasenfratz, P.~Hasenfratz, R.~Lacaze, F.~Niedermayer and M.~Teper
for valuable discussions and H.A.~Kastrup for continuous support.
One of the authors (M.A.S.) would like to thank RWTH Aachen
and HLRZ J\"ulich for a kind hospitality during his visits.
The calculations have been performed on the CRAY Y-MP of HLRZ J\"ulich.

\newpage
\fpsverbosefalse

\begin{figure}
\centerline{%
\fpsxsize=10.3cm 
\fpsbox[47 413 352 814]{f1.ps}}
\vspace{-3mm}
\caption[fig1]{\small The numerical data for the fermion mass $am_F$ on finite
lattices in the U(1) Yukawa model
plotted against $1/y^2$, where $y$
is the Yukawa coupling constant.
The dotted, dashed and full lines are the mean field predictions
on the $L^2$ lattices, $L=16, 32$ and 64, respectively.
The statistical error bars are omitted to make the plot tidy,
they do not exceed the symbol size.
In this figure $\kp= \lm =0$, which means that the U(1) Yukawa model is
identical to the chiral Gross-Neveu model.}
\end{figure}
\begin{figure}
\centerline{%
\fpsxsize=10.3cm 
\fpsbox[47 413 352 814]{f2.ps}}
\vspace{-3mm}
\caption[fig2]{\small As in fig.~1, but now for $\kp = 0.2$ and $\lm = 0$.
Here both the $1/N_F$ expansion and the MFA are as well applicable
as in the GN case.}
\end{figure}
\begin{figure}
\centerline{%
\fpsxsize=10.3cm 
\fpsbox[47 413 352 814]{f3.ps}}
\vspace{-3mm}
\caption[fig3]{\small As in fig.~1, but now for $\kp = 0$ and $\lm = \infty$.
The $1/N_F$ expansion is not applicable but the MFA
still describes the data very well.}
\end{figure}
\begin{figure}
\centerline{%
\fpsxsize=10.3cm 
\fpsbox[47 413 352 814]{f4.ps}}
\vspace{-3mm}
\caption[fig4]{\small As in fig.~1, but now for $\kp = 0.2$ and $\lm = \infty$.
The $1/N_F$ expansion is not applicable and also the MFA
does not describe the data very well.
The dotted-dashed line represents a fit to the $64^2$ data (circles)
by means of
the generalized gap equation \protect{(\eq{GAP})}.}
\end{figure}

\end{document}